\documentclass[conference]{IEEEtran}
\IEEEoverridecommandlockouts
\usepackage{cite}
\usepackage{amsmath,amssymb,amsfonts}
\usepackage{algorithmic}
\usepackage{graphicx}
\usepackage{textcomp}
\usepackage{etoolbox}
\def\BibTeX{{\rm B\kern-.05em{\sc i\kern-.025em b}\kern-.08em
    T\kern-.1667em\lower.7ex\hbox{E}\kern-.125emX}}
\begin{document}

\title{SDN-based In-network Honeypot: Preemptively Disrupt and Mislead Attacks in IoT Networks\\
}

\author{
	\IEEEauthorblockN{Hui Lin}
	\IEEEauthorblockA{\textit{Computer Science and Engineering Department} \\
		\textit{University of Nevada at Reno}\\
		Reno, NV, USA\\
		hlin2@unr.edu}
}

%% to reduce the marging between title box and authors
\makeatletter
\patchcmd{\@maketitle}
  {\addvspace{0.5\baselineskip}\egroup}
  {\addvspace{-1\baselineskip}\egroup}
  {}
  {}
\makeatother

\maketitle

\newenvironment{myitemize}{%
	\edef\backupindent{\the\parindent}%
	\itemize%
	\setlength{\parindent}{\backupindent}%
}{\enditemize}

\begin{abstract}
Detecting cyber attacks in the network environments used by Internet-of-things (IoT) and preventing them from causing physical perturbations play an important role in delivering dependable services. To achieve this goal, we propose in-network Honeypot based on Software-Defined Networking (SDN) to disrupt and mislead adversaries into exposures while they are in an early stage of preparing an attack. Different from traditional Honeypot requiring dedicated hardware setup, the in-network Honeypot directly reroutes traffic from suspicious nodes and intelligently spoofs the network traffic to them by adding misleading information into normal traffic. Preliminary evaluations on real networks demonstrate that the in-network Honeypot can have little impact on the performance of IoT networks.  
\end{abstract}

\begin{IEEEkeywords}
IoT, Software-Defined Networking, Honeypot
\end{IEEEkeywords}

\vspace{-0.5em}
\section{Introduction} \vspace{-0.5em}
Many critical infrastructures, e.g., smart meters and smart homes, have used Internet-of-things (IoT) to monitor and control physical processes. Detecting cyber attacks in the IoT network environment and preventing them from causing physical perturbations play an important role in delivering dependable services. Based on historical incidents in industrial control systems, ``remote insider'' attacks can become a big threat to IoT networks~\cite{lee2016analysis}. After penetrating into the IoT networks, adversaries stay in a preparation stage, during which they use the existing network traffic to study computing context and collect information related to the physical processes. With the help of collected information, adversaries can develop and execute attack concept of operations by crafting malicious commands in legitimate formats without raising anomaly alerts at network levels.

Current detection approaches for IoT networks are passive~\cite{Gendreau2016survey}. Those approaches rely on anomaly patterns of communication networks after adversaries perform malicious activities, e.g., propagating malware, issuing malicious commands~\cite{hodo2016threat}. The communication nodes in an IoT networks are used to operate physical processes, e.g., opening or closing a valve, adjusting room temperature. After adversaries inject malicious activities into IoT networks, the impact of attacks can quickly propagate through underneath physical processes over a wide geographic area~\cite{ronen2017iot}. Consequently, even though some passive detection approaches can identify malicious activities, it is very challenging to prevent physical damage. 

We argue that preemptive approaches, such as Honeypot or Honeynet, which can expose adversaries at an early stage are effective ways to prevent damage caused by attacks. In recent years, multiple projects use traditional Honeypot for cyber-physical industrial control systems, to attract adversaries and trace their  activities~\cite{wilhoitgaspot}\cite{buza2014cryplh}. However, it is challenging to apply traditional Honeypot into IoT networks, due to three reasons. \textit{First}, IoT networks can contain a large number of communication nodes, which makes it challenging to mimic the network of the similar size. \textit{Second}, IoT network has a very dynamic feature; the participating nodes and their connections can experience fast changes. Once a Honeypot is built, it is difficult to update its implementation according to run-time changes. \textit{Third}, traditional Honeypots lack the support for constructing meaningful application-layer payloads, e.g., measurements exchanged between communication nodes in IoT networks. Randomly generated measurements communicated in Honeypot can reveal the presence of a bogus environment to adversaries. 

Compared to traditional Honeypots, we propose in this paper an in-network Honeypot by using traffic-manipulation capability enabled by Software-Defined Networking (SDN). The in-network Honeypot does not require setting up a dedicated network environment; it directly reroutes the traffic from suspicious nodes identified at run time to an SDN controller, which effectively quarantines the suspicious nodes from other communication nodes. The SDN controller spoofs network communications, which are used to interact with suspicious nodes, on behalf of nonexistent nodes, which we refer as \textit{phantom nodes}. There is no dedicated hardware or software resources allocated for the phantom nodes; their existence, e.g., the IP addresses, are only reflected on spoofed packets issued from the SDN controller to the suspicious nodes. Furthermore, we include in the spoofed traffic misleading information, such as vulnerability of certain physical processes to mislead adversaries into targeting on phantom nodes to perform malicious activities. Consequently, the in-network Honeypot can detect adversaries' malicious activities in a quarantined environment without causing real physical disruptions of the protected IoT networks.
\vspace{-0.5em}

\begin{figure*}[!htp]
\centering
  \includegraphics[width=.8\textwidth]{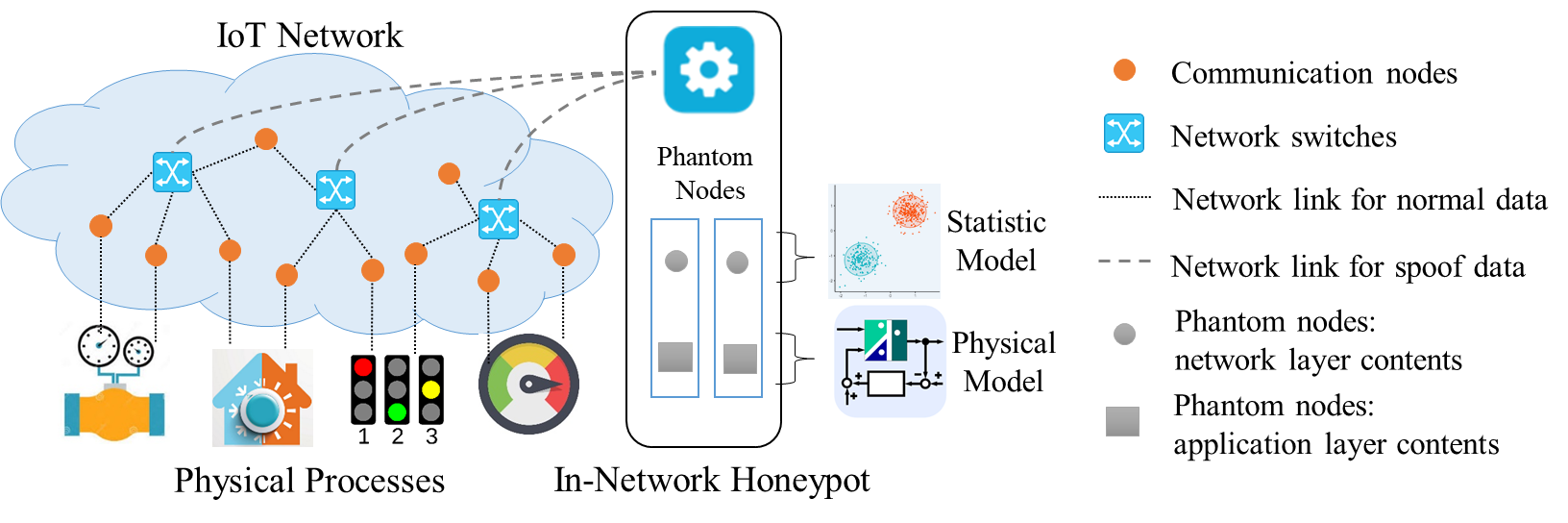}
  \caption{The architecture of in-network Honeypot for IoT.}
  \label{figure-arch}
\end{figure*}
%\vspace{-1em}

\section{System Assumption and Attack Model}\vspace{-0.5em}
In this paper, we consider IoT networks which rely on IP-based networks for communications. As shown in Figure~\ref{figure-arch}, communication nodes exchange information about underneath physical processes, such as thermostats or smart meters. There are two common operations performed in IoT networks: control operations used to configure or operate physical processes and periodic polling operations used to collect analog measurements indicating the state of the physical processes.

% TODO adding reference of penetrations?
We consider the ``remote insider'' threat model in this paper. \textit{First}, we assume that adversaries can penetrate communication nodes in IoT networks from public networks, which make adversaries insiders. To become insiders, adversaries can exploit vulnerabilities in employee's devices, e.g., laptops, smart phones, or USB drives, that are connected to the IoT network. \textit{Second}, we assume that adversaries are ``remote'' (i.e., not an expert) to the configuration of IoT networks as well as the characteristics of underneath physical processes. Under these assumptions, adversaries can monitor information exchanged over IoT networks and thus obtain the knowledge on physical processes to prepare for malicious operations. 

We assume that the IoT networks under protections can support SDN-enabled switches and the functionality of SDN controlling plane is trusted. SDN is a new network paradigm whose key feature is the separation of the control plane and the data plane~\cite{albert05ccr}. In SDN, network switches are simple forwarding devices, whose forwarding rules can be dynamically configured by a central controller. In recent years, many telecommunication companies, e.g., Huawei, began integrating SDN into their core wireless networks~\cite{huaweisdn}. In the survey shown in~\cite{bera2017software}, Bera et al. has presented the opportunities for SDN to address critical challenges in IoT, e.g., dynamic network management, resource allocations, and resilience. Even though SDN can facilitate the implementation of the in-network Honeypot, the concept of spoofing network traffic to disrupt and mislead adversaries is not restricted by SDN but can be implemented by any traffic manipulation techniques.        

%any communication path that connects the control center and end devices. Those computing devices, such as human-machine interfaces (HMI) or RTUs, are often installed at substations located over a large geographical area. Because it is challenging to maintain computing devices in a wide area, we often find unpatched vulnerabilities in those devices, e.g., an old TCP vulnerability found in substation devices [8]. In most well-known attacks targeting industrial control systems, e.g., Stuxnet and attacks against Ukrainian power system [1][2], adversaries targeted vulnerable computing devices and used them as footholds to prepare and execute attacks 

\section{Architecture of In-network Honeypot} \vspace{-0.5em}
%TODO forward references
We present the overall design of the in-network Honeypot in Figure~\ref{figure-arch}, whose implementation relies on SDN. An SDN controller can observe all communication going through network switches under its control and use the global knowledge of a communication network to make a traffic-management decision that can achieve optimal network performance, resource utilization, or reliability. The proposed in-network Honeypot, however, is to use the knowledge obtained by the SDN controller and its programmable capability to achieve two objectives: (1) quarantine suspicious or potential malicious nodes and (2) mislead adversaries targeting on nonexistent communication nodes (i.e., phantom nodes). Achieving these two objectives, we can detect adversaries while they are preparing attacks and prevent them from causing damage to physical processes.   

%\noindent
%\textit{\textbf{Quarantining suspicious nodes.}} 
\vspace{-5pt}
\subsection{Quarantine suspicious nodes}

We assume that existing IoT networks have intrusion detection systems (IDS) to raise alerts on suspicious activities from communication nodes. After IDS identifies suspicious nodes, the in-network Honeypot uses SDN controllers to stop forwarding the traffic from the suspicious nodes to other communication nodes. In other words, the network traffic from the suspicious nodes cannot reach any communication nodes, which make them quarantined from the IoT networks. Regarding all network traffic from the suspicious nodes, the SDN controller responds with spoofed information on behalf of the destination nodes. Consequently, the suspicious nodes communicate with phantom nodes spoofed by the SDN controller with no attachment to physical machines or software processes.    

%\noindent
%\textit{\textbf{Spoofing communication.}} 
\vspace{-0.4em}
\subsection{Spoof communication}

It is critical for the in-network Honeypot to spoof communication that can mimic the real network traffic, which can maintain highly active interactions between phantom nodes and suspicious nodes. Based on these interactions, the in-network Honeypot can collect more information about those suspicious nodes to make a trustworthy decision. 

We spoof communications by adding variations into normal communication patterns. In addition, we include some misleading information in the variations to mislead adversaries into targeting phantom nodes instead of real communication nodes. Note that an SDN controller can observe all traffic going through the switches under its control. We can integrate into the SDN controller anomaly-based intrusion detection techniques to build a normal pattern of each communication node. 

As shown in Figure~\ref{figure-arch}, the in-network Honeypot needs to construct contents at both network layer and application layer to spoof network packets on behalf of phantom nodes. We propose using different approaches to construct contents, i.e., using statistic models for network layer contents and using physical models for application layer contents. 

\noindent
\textit{\textbf{Construct contents at network layer}}

To construct contents at network layers, we first build statistic models that can classify communication nodes based on their network-layer characteristics. The example of these characteristics includes the range of IP-addresses, the length of network packets, the latency between different types of network packets. Because many physical processes associated with communication nodes run fixed operations and follow deterministic patterns, the in-network Honeypot can use network traffic observed at the SDN controller to fingerprint physical processes and classify network communications associated with them~\cite{formby2016s}. The advantage of using SDN controllers is that the in-network Honeypot does not need to interfere the normal physical processes but to rely on observed network traffic to build statistic models. Then at runtime, the in-network Honeypot creates contents by following these statistic models. 

% adding something about adversarial learning

\noindent
\textit{\textbf{Construct contents at application layer}}

In order to prevent adversaries from disrupting underneath physical processes, we construct contents at the application layer to (1) mimic the state of physical processes and (2) mislead adversaries into disrupting the nonexistent physical processes controlled by phantom nodes.

\vspace{-0.5em}
\begin{figure}[!htp]
\centering
  \includegraphics[width=.35\textwidth]{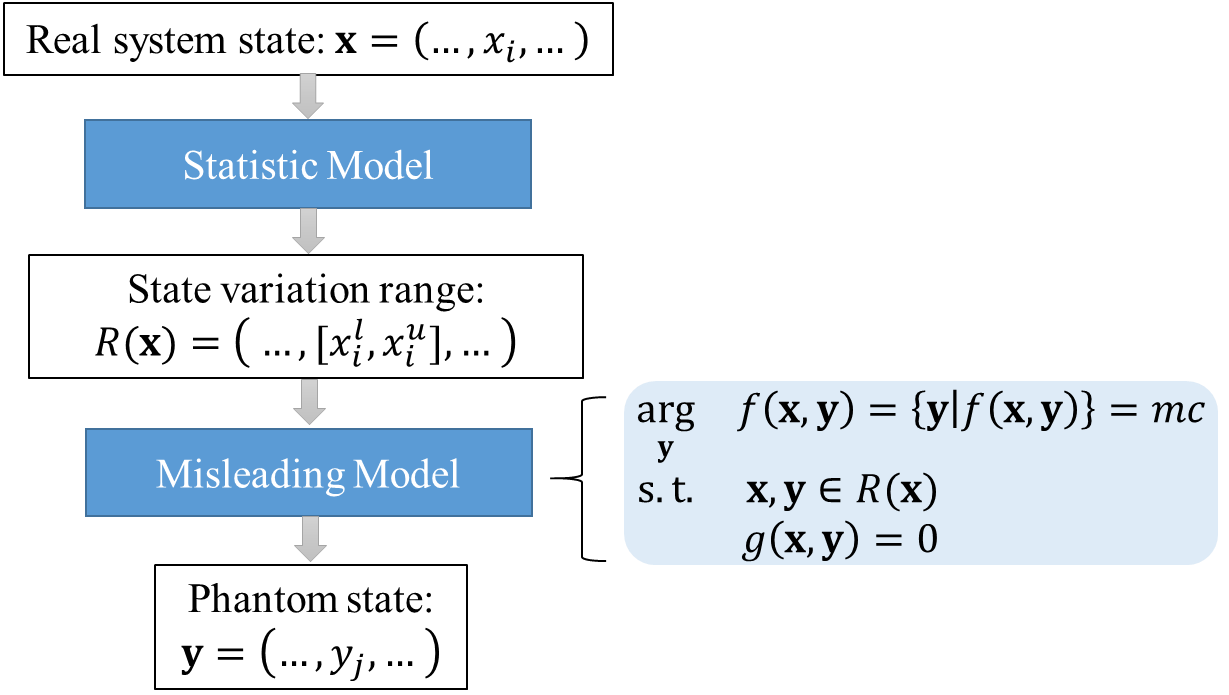}
  \caption{Procedure to construct contents at application layer.}
  \label{figure-procedure}
\end{figure}
\vspace{-0.5em}

In Figure~\ref{figure-procedure}, we present the high-level procedure to construct contents at the application layer. We use a vector \textbf{\textit{x}} to represent the state of physical processes managed by the IoT network. Through the traffic collected by the SDN controller, we first monitor those state variables and model their statistic characteristics (in ``statistic model''), such as the variation range of each state determined by its lower bound and upper bound (i.e., $x_i^l$ and $x_i^u$ in the figure). In the next step, we include the state variation range in an optimization problem in the ``misleading model.'' The solution to this problem is ``phantom state'', which represents the state of nonexistent physical processes that are controlled by phantom nodes. In other words, \textit{phantom states are the contents at the application layer of the network packets issued from the phantom nodes}. 

Assume that we use function $f(x,y)$ to represent decision procedure of adversaries to cause physical damage, i.e., $mc$, based on the observed system state and phantom state. In the ``misleading model,'' we set $mc$ as the operations that can cause significant disruption on the phantom state but little impact on real physical processes. Consequently, the solutions to the optimization problem determine the phantom states that can mislead adversaries into designing ineffective attack strategies, which introduce no physical damage. Note that the optimization makes both physical and phantom states subject to the constraints of state variation range observed at runtime. This constraint makes the resulting phantom state follow normal variations and avoid adversaries' suspicions. Also, because physical processes need to follow physical laws, we add the constraint to make physical and phantom state consistent with the physical laws in a mathematical expression, i.e., $g(x,y)=0$ in the figure. Because this optimization problem mimics adversaries' decision procedure, solving the problem requires similar computation complexity as required to prepare attacks.   % TODO add examples? 

%The instantiation of the optimization problem varies according the physical processes. For example, in a smart grid, the malicious operation can be a command to disconnect relays to cause large power imbalance. But in a water treatment plant, the malicious operation can be the delay of a command that opens a output port of a tank to avoid water floods.    
\vspace{-0.5em}
\subsection{Handle false detection}
If IDS makes a false positive detection and a suspicious node is mistakenly quarantined, the in-network Honeypot can use the SDN controller to restore its communication path. Additionally, the SDN controller can profile the physical changes initiated by the suspicious nodes. After restoring communications for suspicious nodes, the in-network Honeypot can use the profile as reference points to update the physical process, to avoid repeated operations from the suspicious nodes. To profile the physical changes initiated by the suspicious nodes at runtime and without causing real physical changes, we can use simulations, which represent the mathematical models of the physical process, to estimate the consequences of the commands and record them.

In addition to handling the false positive detection after their occurrence, the in-network Honeypot can reduce the number of false detection in advance by increasing the accuracy of anomaly-based IDS. Anomaly-based IDSes raise an alert when they observe any network traffic that deviates from normal patterns. They suffer from two drawbacks: (1) raising false positive alerts on anomaly not due to attacks and (2) introducing false negative detection if the IDSes build the normal patterns based on the network traffic that has already been contaminated by malicious activities~\cite{sommer2010outside}. With the help of the in-network Honeypot, we can remedy the negative impacts caused by these two drawbacks. When the in-network Honeypot identifies a suspicious node as malicious, it can use the interactions with them to build the model of adversaries in parallel with the model of normal traffic. The adversaries' model can help to (1) reduce false positive detection by distinguishing attacks from anomalies and (2) reduce false negative detection by removing malicious traffic of suspicious nodes from the model to build the normal traffic patterns.

%\begin{myitemize}
%\item 
%\textit{Reduce the number of false positive detection.}  One of the major problems of anomaly-based IDS is that it only builds the pattern of normal traffic without any information regarding to malicious traffic. If the suspicious nodes turn out to be real malicious nodes, the in-network Honeypot catches rare and precious malicious traffic, which can be further used to refine the statistic model built in the IDS and accurately classify attacks. 

%\item 
%\textit{Reduce the number of false negative detection.} When the in-network Honeypot identifies a suspicious node as malicious, it is very likely that the network traffic of this node is used in anomaly-based IDS to build normal communication pattern. Based on the decision of suspicious nodes, IDSes filter out malicious traffic of this node from the model to build the normal traffic patterns.

%\end{myitemize}

\section{Evaluation}\vspace{-0.5em}
%\noindent
%\textbf{\textit{Environment.}} 
\subsection{Environment}
\begin{figure}[!htp]
\centering
  \includegraphics[width=.45\textwidth]{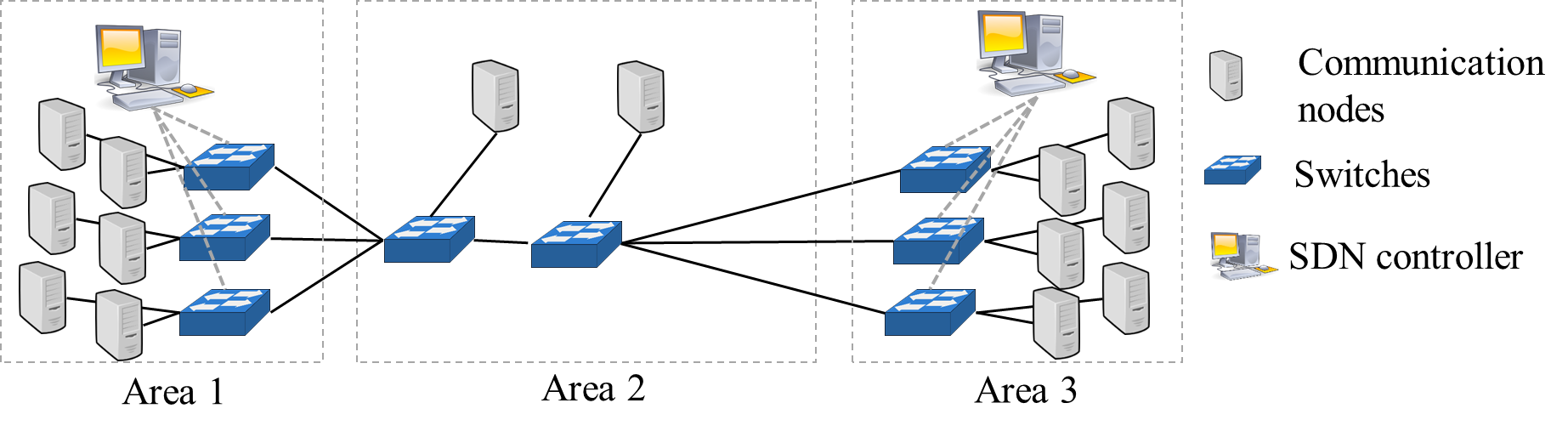}
  \caption{Dumbbell topology to simulate IoT networks.}
  \label{figure-dumbbell}
\end{figure}

We used the GENI testbed, a nationwide network experiment platform, to construct IoT networks with a dumbbell topology shown in Figure~\ref{figure-dumbbell}~\cite{geni}. In the GENI testbed, we used real SDN-enabled hardware switches to connect virtual machines that simulate communication nodes. Because in GENI testbed we can use virtual machines physically located in three different areas, the evaluations can reflect the performance of wide area communications in IoT networks. We also constructed networks of three different sizes by changing the number of communication nodes connected to switches. When indicating a network, we add the number of nodes with the name of the network topology in parentheses.

We used DNP3 as the network protocol to exchange information~\cite{dnp3protocol}. Although the DNP3 protocol is mainly used in electric and water companies, its complex structure and rich data formats can cover wide varieties of measurements and operations used in different IoT networks.  

\vspace{-0.5em}
\vspace{-0.5em}

We implemented the in-network Honeypot in ONOS, an open-source network operating systems commonly used as SDN controllers in commercial networks~\cite{berde2014onos}. Because we used DNP3 as the protocol for communications, we included in ONOS an encoder to spoof DNP3 packets.  

\subsection{Evaluation results}

\begin{figure}[!htp]
\centering
  \includegraphics[width=.4\textwidth]{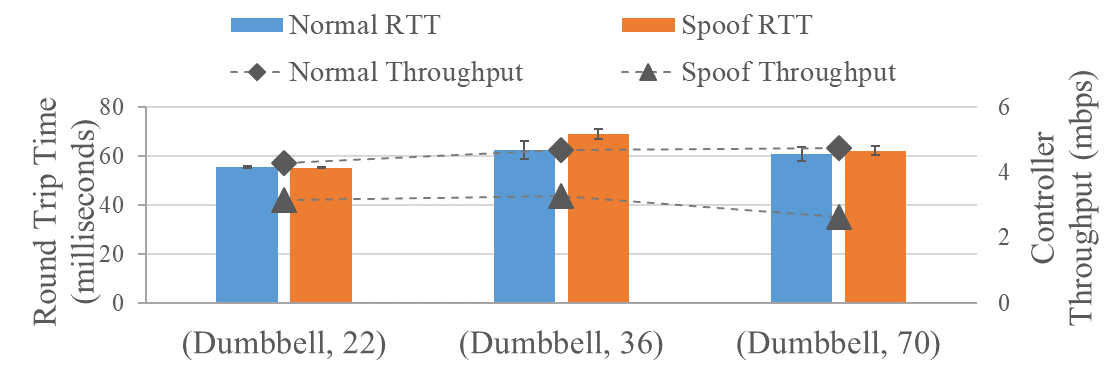}
  \caption{Performance evaluations on spoofing network traffic (with 99\% confidence interval).}
  \label{figure-performance}
\end{figure}

%To demonstrate the feasibility of applying in-network Honeypot in IoT networks, 
In this section, we evaluate the impact of spoofing network traffic on the performance of IoT networks. Specifically, we spoofed the traffic of 80\% computing nodes to simulate a worst case scenario. As shown in Figure~\ref{figure-performance}, we use two performance metrics for evaluations, i.e., the round trip times (RTT) and the throughput of the SDN controller. 

% TODO: change the marker
\vspace{-0.5em}
\vspace{-0.5em}

On the primary \textit{y}-axis, we compare the RTT when networks uses the in-network Honeypot to spoof traffic (``spoof RTT'') to ``normal RTT.'' From the figure, we can observe that the change of RTT is within 10\% (less than 10 milliseconds). Consequently, it can be challenging for adversaries to distinguish the spoofed traffic from the real one based on the observed variations in RTTs. 

On the secondary \textit{y}-axis, we compare the throughput of the SDN controller spoofing network traffic (``spoof throughput'') with the throughput when it directly forwards network traffic (``normal throughput''). Compared to the ``normal throughput,'' we can see that there was an approximately 30\% decrease on average; throughput of spoofing network packets varied between 2.8 Mbps and 3.5 Mbps. For a DNP3 packet of 1 kilobyte (KB), which can contain more than 200 32-bit measurements, the SDN controller can spoof more than 300 packets per second. 

\section{Conclusions} \vspace{-0.5em}
In this paper, we propose an in-network Honeypot, which reroutes network traffic from suspicious nodes to an SDN controller to quarantine their communications in an IoT network. After quarantining the suspicious nodes, the SDN controller spoofs network communication with suspicious nodes on behalf of nonexistent phantom nodes. We use both statistic model and physical model to construct contents of the spoofed packets. The spoofed packets can mislead adversaries into targeting phantom nodes and thus to prevent potential physical damage from happening on real communication nodes and the underneath physical processes. Preliminary evaluations on real SDN networks demonstrate that deploying in-network Honeypot introduces small overhead on normal network communications. 
 
\bibliographystyle{IEEEtran}
%\bibliography{IEEEabrv}
\bibliography{InnetworkHoneypotR}

\end{document}